\documentstyle[12pt,agums]{article}

\lefthead{CHANG ET AL.}

\righthead{Coherent Magnetic Structures and BBF}

\paperid{}

\cpright{AGU}{1999}

\ccc{0148-0227/97/97GL-00000\$05.00}

\authoraddr{Tom Chang, Center for Space Research,
 Massachusetts Institute of Technology, Cambridge, Massachusetts 02139.}

\authoraddr{Cheng-chin Wu, 
Department of Physics and Astronomy,
University of California, Los Angeles,
California 90095.}

\authoraddr{Vassilis Angelopoulos, 
Space Sciences Laboratory,
University of California, Berkeley,
California 94720.}

\begin{document}

%
%



\title{Preferential Acceleration of Coherent Magnetic Structures and
Bursty Bulk Flows in Earth's Magnetotail}


%
%

\author{Tom Chang}
\affil{Center for Space Research, Massachusetts Institute of Technology,
 Cambridge, Massachusetts}

\author{Cheng-chin Wu}
\affil{Department of Physics and Astronomy, University of California, 
Los Angeles, California}

\author{Vassilis Angelopoulos}
\affil{Space Sciences Laboratory,
University of California, Berkeley, California}

\begin{abstract}
Observations indicate that the magnetotail convection is turbulent and bi-modal, 
consisting of fast bursty bulk flows (BBF) and a nearly stagnant background. We 
demonstrate that this observed phenomenon may be understood in terms of the 
intermittent interactions, dynamic mergings and preferential accelerations of 
coherent magnetic structures under the influence of a background magnetic field 
geometry that is consistent with the development of an X-point mean-field 
structure.

\end{abstract}

\begin{article}
\section{Introduction}
Recent satellite observations indicate that the Earth's magnetotail is generally 
in a state of intermittent turbulence constantly intermixed with localized fast 
bursty bulk flows (BBF) [{\it Angelopoulos et al.}, 1996; {\it Lui}, 1998].  A 
model of sporadic and localized merging of coherent magnetic structures has been 
proposed by {\it Chang} [1999; and references contained therein] to describe the 
dynamics of the magnetotail. When conditions are favorable, the coherent 
structures may merge, interact, convect and evolve into forced and/or self-
organized critical (FSOC) states [{\it Chang}, 1992], a condition satisfied by the 
observational statistics of the burst duration characteristics of the BBF [{\it 
Angelopoulos}, 1999]. When considered with the realistic magnetotail geometry, 
the coarse-grained dissipation of turbulent fluctuations produced by the 
interacting multi-scale coherent structures may trigger ``fluctuation-induced 
nonlinear instabilities'' and possibly initiate substorms. Some of the above 
ideas have been demonstrated by direct numerical simulations
 [{\it Wu and Chang}, 
2000]. In this paper, we address the dynamics, particularly the preferential 
acceleration, of the coherent structures in sheared magnetic field geometry. 
Preliminary comparisons of the theoretical ideas and new numerical results with 
the observed characteristics of BBF will also be presented.



\section{Observational Properties of BBF}

In situ observations [{\it Angelopoulos et al}., 1999; {\it Nagai et al.}, 1998 
and references contained therein] indicate that BBF are an important means of 
magnetotail transport.  Removing the fast BBF, the remaining flow state has a 
small average convection.  The BBF are seen during precursor activity, at 
substorm expansion and recovery phases.  

  Near-Earth BBF are primarily Earthward and consist of multiple, 
short-lived flow bursts, i.e., they are intermittent. 
Distribution functions of mid-tail BBF near the source region 
of a tailward-retreating reconnection site at late substorm 
expansion phase show that multiple, localized acceleration 
sites are responsible for the observed flows 
[{\it Angelopoulos et al}., 1996]. 
Distant tail fast flows are tailward, 
they are associated with plasmoids, they 
also have multiple flow peaks and multiple current 
filaments within them, and they are likely localized in 
the cross-tail dimension (e.g. {\it Ieda et al}. [2001] 
and references therein). Thus although preferential acceleration 
is observed on either side of a mid-tail region, probably associated 
with the gross formation of an X-point topology, 
the flows and the currents within them are rather intermittent.

When considered in connection with an 
observed X-point, the primary directions of convection of BBF are earthward 
(tailward) when they are situated earthward (tailward) of the X-point [{\it 
Nagai et al.}, 1998].  It is therefore of preponderate interest to decipher the 
fundamental physical mechanism that are responsible for the preferential 
acceleration and convective nature of the observed BBF.

It had been suggested by {\it Chang} [1992] that the magnetotail could exhibit 
dynamic forced and/or self-organized critical (FSOC) behavior. Statistical studies of the AE index [{\it Consolini}, 1997] and global auroral imagery [{\it Lui et al.}, 2000] during substorm times seem to indicate consistency with such a dynamic description. More recently, {\it 
Angelopoulos et al.} [1999] demonstrated specifically that the probability 
distribution of the BBF duration could also be described by a power law, an 
indication of dynamic criticality.  A physically realistic topological model 
based on the coherent magnetic structures was proposed by {\it Chang} [1998, 
1999] to explain the multiscale nature of the intermittent turbulence that can 
lead to the FSOC state in the magnetotail and its low dimensional behavior [{\it Klimas et al.}, 1996].
It was also suggested that the 
coarse-grained dissipation of the fluctuations associated with FSOC could lead 
to nonlinear instabilities and trigger the development of mean-field X-point 
magnetic structures in sheared magnetic fields. We shall demonstrate below that 
these ideas lead naturally to the preferential acceleration and observed 
convective nature of the fast BBF.

\section{Coherent Magnetic Structures in Earth's Magnetotail}
There is considerable evidence that many turbulent flows are far 
from being totally disorganized. Indeed they can posses 
``coherent structures,'' as indicated by experiments and 
numerical simulations [e.g., {\it Frisch}, 1995]. {\it Chang} [1998] 
suggested that the coherent structures are basically current 
filaments in the neutral sheet region of the magnetotail.

Most field theoretical discussions begin with the concept of 
propagation of waves. For example, in the MHD formulation,  one 
can combine the basic equations and express them in the following 
propagation forms:$$\rho d{\bf V}/dt={\bf B}\cdot \nabla{\bf  
B}+\cdot \cdot \cdot\eqno(1)$$
$$d{\bf B}/dt={\bf B}\cdot \nabla{\bf V}+\cdot \cdot \cdot \eqno(2)$$
where the ellipsis represents the effects of the anisotropic 
pressure tensor, the  compressible and dissipative effects, and 
all notations are standard. Equations (1,2) admit the well-known 
Alfv\'en waves. For such waves to propagate, the propagation 
vector ${\bf k}$ must contain a field-aligned component, i.e., 
${\bf B} \cdot \nabla \rightarrow i{\bf k} \cdot {\bf B} \ne 0$. 
However, near the singularities where the parallel component of 
the propagation vector vanishes (the resonance sites), the 
fluctuations are localized. That is, around these resonance 
sites (usually in the form of curves in physical space), it may 
be shown that the fluctuations are held back by the background 
magnetic field, forming coherent structures in the form of flux 
tubes [{\it Chang,} 1998; 1999].  For the neutral sheet region of the magnetotail, these coherent structures are 
essentially force-free and in the form of current filaments in the 
cross-tail direction [{\it Chang}, 1998].
If the cross-tall current is primarily carried by the coherent 
magnetic structures, these structures will be strongly affected 
by the magnetic field due to the Lorentz force.  As they are 
being preferentially accelerated they will dynamically deform and 
dissipate fluctuations to produce smaller structures, while 
occasionally interact and merge with neighboring structures.

\section{Preferential Acceleration of BBF}
Consider the coherent magnetic structures discussed in the previous 
section, which
in the neutral sheet region of the plasma sheet are essentially filaments of 
concentrated
currents in the cross-tail direction.  As we have argued previously, in a 
sheared magnetic field, these multiscale fluctuations can supply the required 
coarse-graining dissipation that can produce nonlinear instabilities leading 
to X-point-like structures of the average magnetic field lines.  For a 
sheared magnetic field $B_x(z)$, upon the onset of such fluctuation-induced nonlinear 
instabilities, the average magnetic field will 
generally acquire a $z$-component. 
Let us choose $x$ as the Earth-magnetotail direction (positive 
toward the Earth) and $y$ in the cross-tail current direction.  Then the 
deformed magnetic field geometry after the development of an X-point 
structure will generally have a positive (negative) $B_z$ component earthward 
(tailward) of the X-point.  Near the neutral sheet region, the Lorentz force 
will therefore preferentially accelerate the coherent structures  earthward if 
they are situated earthward of the X-point and tailward if they are situated 
tailward of the X-point, Figure 1.  These results would therefore match 
the general directions of motion of the observed BBF in the magnetotail [{\it 
Nagai et al.}, 1998].

We have performed two-dimensional numerical simulations to verify these
conjectures.  The simulations are based on a compressible MHD 
model that has been used by {\it Wu and Chang} [2000] in previously studies 
of coherent magnetic structures.
In the first example we started the numerical simulation 
initially consisting of randomly distributed magnetic 
fluctuations with zero mean magnetic field. 
The calculation was carried out with 256 by 256 
grid points in a doubly periodic ($x$,
$z$) domain of length $2 \pi$ in both directions.   Eventually, 
after some elapsed time, multiscale coherent magnetic structures 
are formed, which are shown in Fig. 2(a) for $1.5\pi \le x \le 
2\pi$ and $1.1\pi \le z \le
1.6\pi$. The maximum magnitude of the magnetic field is 0.014. 
In this and following examples,
both
pressure and density are about 1; the system has a large plasma 
$\beta$ to
mimic the current sheet region. 
 At this time, a uniform 
$B_z=0.002$ is
introduced. In the middle panels of Fig. 2, the effect of the 
Lorentz
force is clearly seen through the motions of the coherent 
structures after
an additional elapsed time of 100. (For all numerical examples,
the unit of time was based 
on the wave speeds. For instance, the sound speed is about 1.3; 
thus a sound wave would traverse a distance of $\pi$ in a time of 2.4.)
If the direction of $B_z$ is reversed,
the motions of the coherent structures are also reversed as seen 
in the
bottom panels of Fig. 2.

In the second example, we have considered the motions of the 
coherent
magnetic structures that developed in a sheared mean magnetic field 
upon the
initial introduction of random magnetic fluctuations. The 
calculation was
carried out with $1280$ by $256$ grid points in a doubly periodic 
($x$,
$z$) domain of length $10 \pi$ in the $x$-direction and $2 \pi$ 
in the
$z$-direction with a sheared magnetic mean field $B_x=0.025 \cos(z)$. 
The
system again has a high plasma $\beta$ with both pressure and 
density
about 1. These structures were generally aligned in the 
$x$-direction near
the neutral line, $z=1.5\pi$, after some elapsed time. A positive
$B_z=0.001$ was then applied. It can be seen (Figure 3) that, 
after an
additional elapsed time of 50, the coherent structures (mostly 
oriented by
currents in the positive $y$-direction), are generally 
accelerated in the
positive $x$-direction; with one exception where a pair of 
magnetic
structures with oppositely directed currents effectively canceled 
out the
net effect of the Lorentz force on these structures. The 
acceleration
continues in time. We have plotted the $x$-component of the flow
velocities, $v_x$, due to the cumulative effect of the Lorentz 
(and
pressure) forces acting on the flow (and in particular the 
coherent
structures) after additional duration of 400 has elapsed (Fig. 
4). We note
that there are a number of peaks and valleys in the 3-dimensional 
display,
with peak velocities of about 0.02, nearly approaching that of 
the
Alfv\'en speed (about 0.025 based on $|B|=0.025$) mimicking the 
fast BBF
that were observed in the magnetotail.
It is to be noted that an individual BBF event may be composed of one, two, or several coherent structures.

In the final plot (Fig. 5), we demonstrate in a self-consistent 
picture 
what might occur near the neutral sheet region of the 
magnetotail. We
injected randomly distributed magnetic fluctuations in a
predominantly sheared magnetic field, prescribed by $B_x=0.1 
\cos(z)$ and
$B_z=0.002 \sin(x/5)$. 
 The calculation was again carried
out with $1280$ by $256$ grid points in a doubly periodic ($x$,
$z$) domain of length $10 \pi$ in the $x$-direction and $2 \pi$ 
in the
$z$-direction. Eventually, a large-scale
X-point like mean field magnetic structure is produced. The 
coherent
structures (aligned in the neutral sheet region) are subsequently
accelerated away from the X-point in both the positive and 
negative
$x$-directions. Contour plots of $v_x$ after some elapsed time 
clearly
indicated such effects.

\section{Summary}

We demonstrated that dynamical evolutions of coherent magnetic structures in an 
initially sheared magnetic field might provide a convenient description of the 
observed fast BBF in the Earth's magnetotail, particularly near the neutral sheet region. In the magnetotail, the coherent structures are essentially current filaments in the cross-tail direction.  
In an X-point mean magnetic field geometry, selective acceleration of these
coherent structures can be accomplished by the Lorentz force.
Simulation results 
demonstrate that the coherent magnetic structures can indeed be preferentially 
accelerated this way near the neutral sheet and in the appropriate directions.

Although the X-point mean magnetic field can be formed by nonlinear 
instabilities due to the stochastic 
behavior of the coherent magnetic structures, it is not necessary for 
the X-point to occur this way.
An X-point magnetic field geometry formed by 
forced magnetic reconnection is likely to
cause preferential acceleration of
 the coherent structures in a similar manner.

Our work is just at the beginning. 
 In the future, we shall consider more general 
initial two-dimensional mean magnetic field geometries.  Moreover, the coherent 
magnetic structures and interactions are
 usually three-dimensional, characterized 
by sporadic and localized current disruptions
 [{\it Lui}, 1996], and the dynamics 
of the coherent structures, intermittent turbulence,
 and global interactions are 
very complicated.  We shall look into such complexities. 
 The characteristics of 
the ion velocity distributions of the observed BBF generally
 have a crescent-moon 
shape  in the $x-z$  plane [{\it Nagai et al.}, 1998].
The localized merging of the coherent structures 
   aligned in the neutral sheet region would produce
    particle jets at the merging point normal to the
    neutral sheet both in the north and south directions
    with velocities of the order of the Alfv\'en speed.
    We suggest that the combined consequence of such
     localized merging and the preferential acceleration
     of the coherent structures would produce ion
     distributions in the BBF of the form as observed.
 We shall begin a study of such detailed kinetic effects.

%
%

\acknowledgments
The work of TSC was partially supported by
grants from AFOSR,  NSF and                  
NASA, and CCW and AV were partially supported by 
NASA grants NAG5-9111 and NAG5-9626, respectively.




%
%

   %
   %
   %

{}
\end{article}

\newpage


%
%



%
%
\begin{figure}
\caption{Top panel: Schematic of a mean field X-point magnetic 
 geometry.  Bottom panel: Heavy arrows depicting the directions of the Lorentz 
 (${\bf J} \times {\bf B}$) forces acting on two typical coherent magnetic 
 structures (cross-tail current filaments) near the neutral sheet.}
\end{figure}

\vskip 6pt
\begin{figure}
\caption{Effect of the Lorentz force on the motion of the coherent
structures. The magnetic fields are shown in the left three panels 
and the flow
vectors are plotted in the right
three panels. All three plots for the magnetic fields have the same scale, with
${\rm max}(|B|)=0.014$ in (a). The flow-vector plots also have the same
scale, with ${\rm max}(|v|)=0.0015$, 0.0026, and 0.025 in (b), (d), and (e),
respectively. The top panels show the formation of multiscale coherent
magnetic structures. At this time,  a uniform $B_z$ is introduced. The
middle panels show the result with $B_z=0.002$ while the bottom plots
provide the result with $B_z=-0.002$, both after an additional elapsed
time of 100.}
\end{figure} 
\vskip 6pt
 
\begin{figure}
\caption{Effect of the Lorentz force on the motions of the 
coherent structures in a sheared magnetic field due to the 
application of a uniform magnetic field $B_z$. The magnetic fields
(a) and flow velocities (b) are plotted in a domain 
$6\pi \le x \le 8\pi$ and $\pi \le
z \le 2\pi$. The maximum velocity in (b) is 0.006.}
\end{figure}
\vskip 6pt


\begin{figure}
\caption{ A 3D perspective plot of $v_x$. Its peak velocities are about
0.02, nearly approaching that of the Alfv\'en speed (about 0.025 based on
$|B|=0.025$). The maximum velocity is about 3 times higher than that at
earlier time shown in Fig. 3.
Note that, because of the difference of scales 
in $x$ and $z$ in the plot, the     
 accelerated structures have
 the over-emphasized sheet-like shapes.
}
\end{figure}
\vskip 6pt
 
 
\begin{figure}
\caption{ A contour plot of $v_x$ in an X-point mean field
geometry. There are 21 contour levels, with a contour interval of 0.0045.
The  highest level is 0.045 near $(x, z) \sim (7.5\pi, 1.5 \pi)$ and the lowest level is -0.045 near $(x, z) \sim (3\pi, 1.5 \pi)$.
The periodic boundary condition forces $v_x \sim 0$ near the
boundary at $x \sim 0$ and $10\pi$.}

\end{figure}

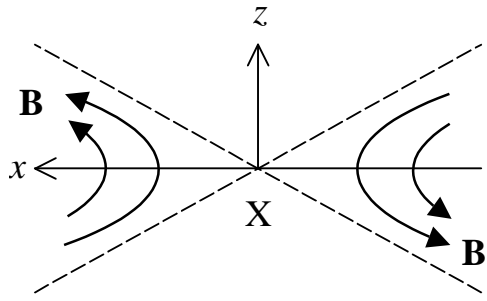
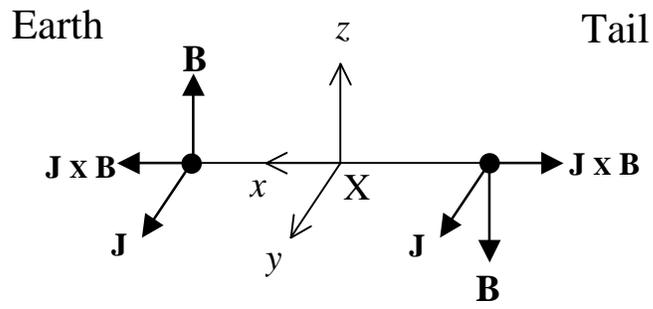
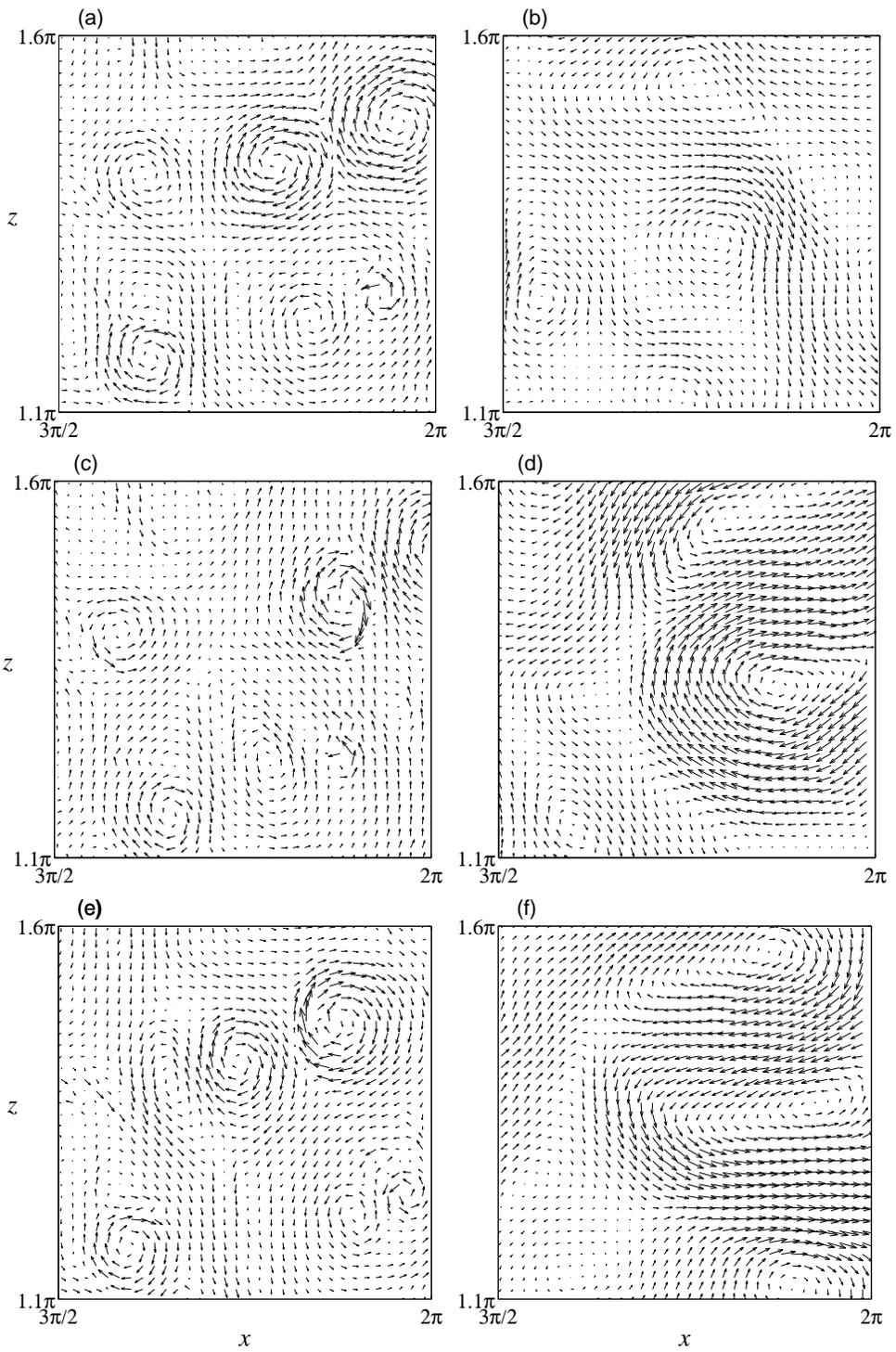
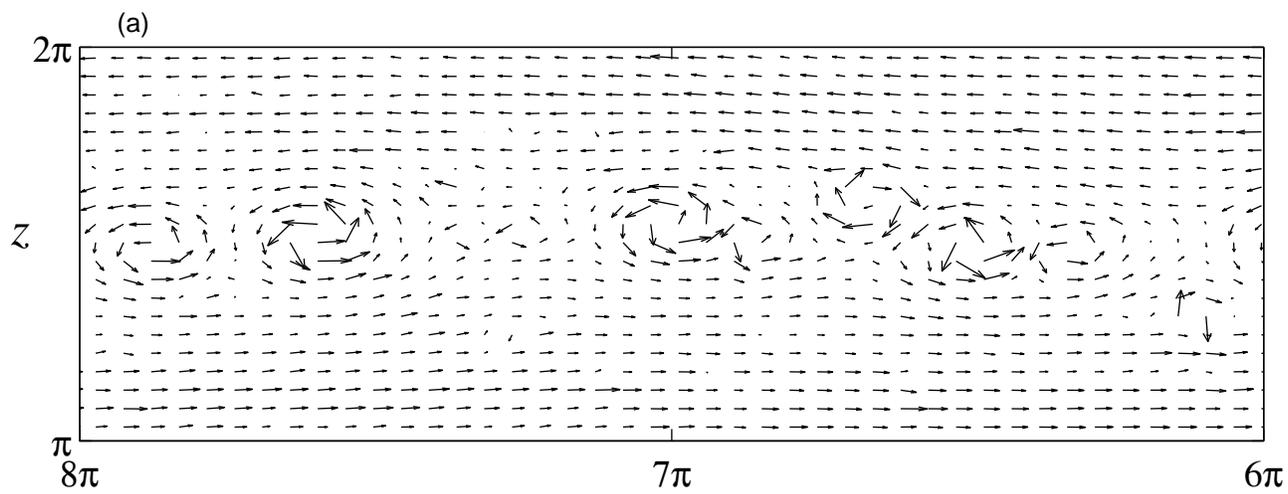
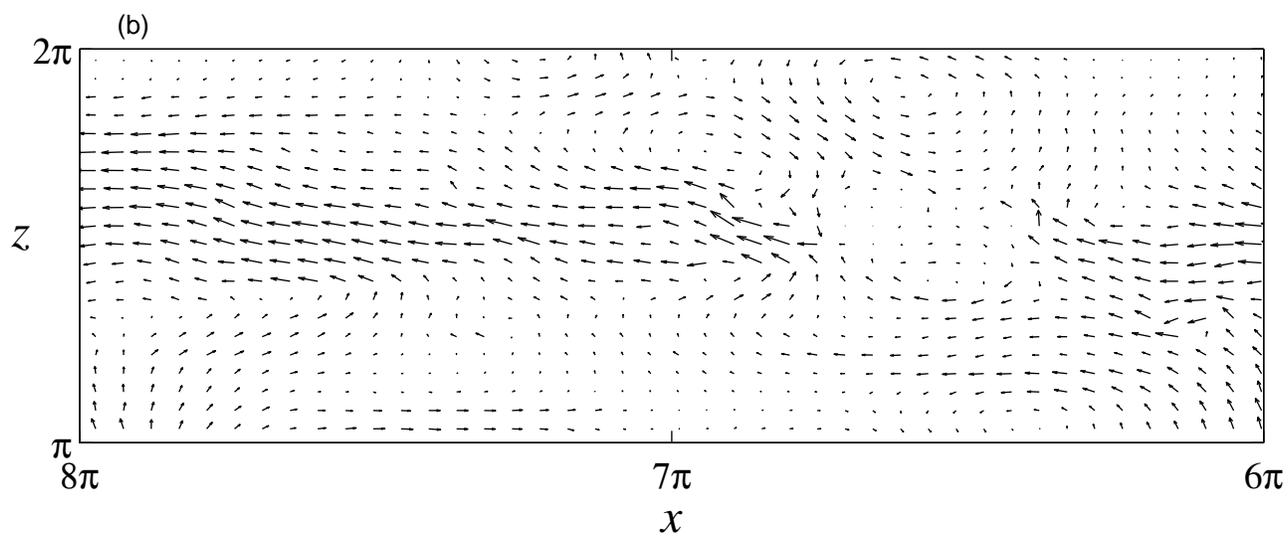
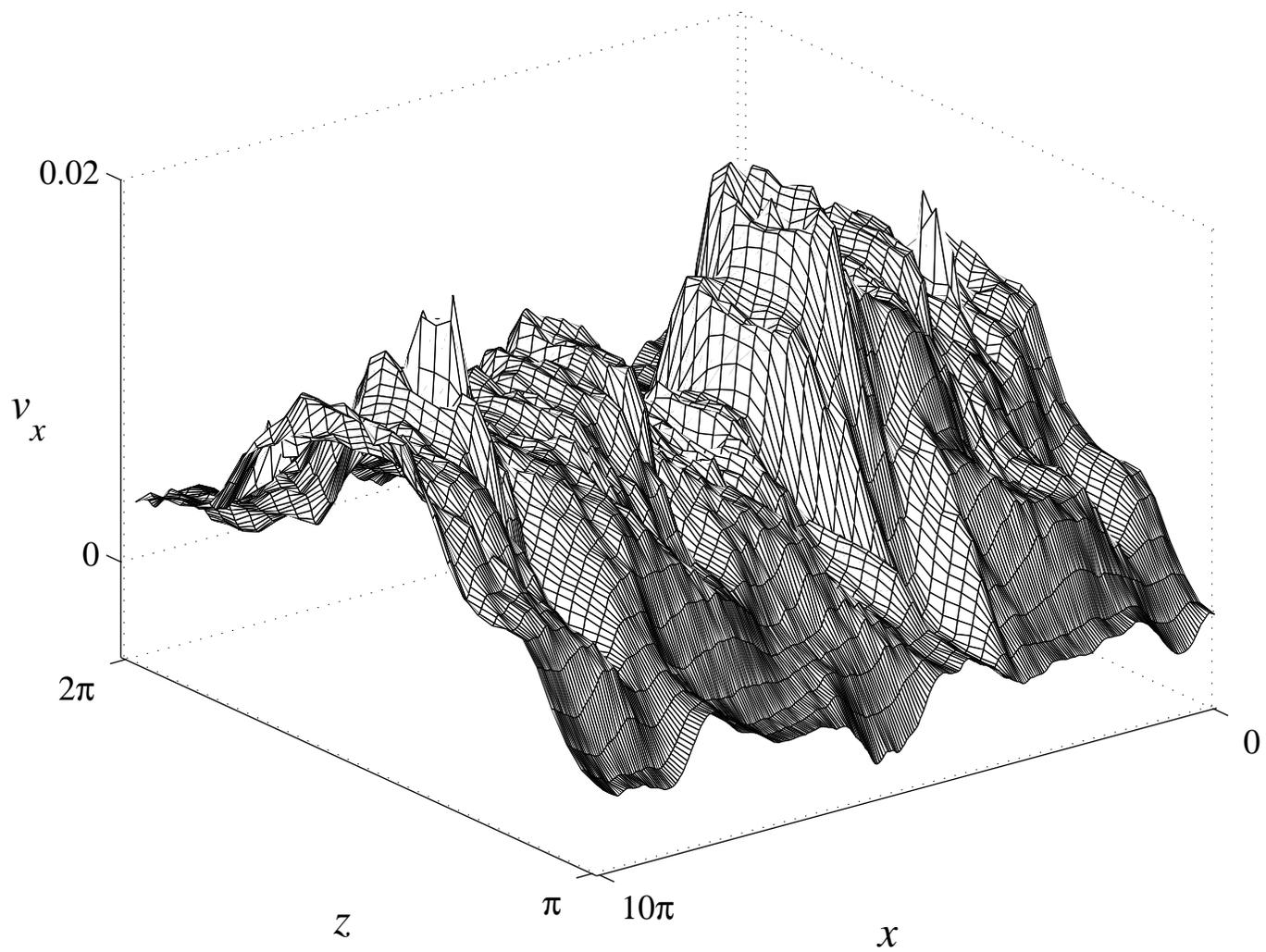
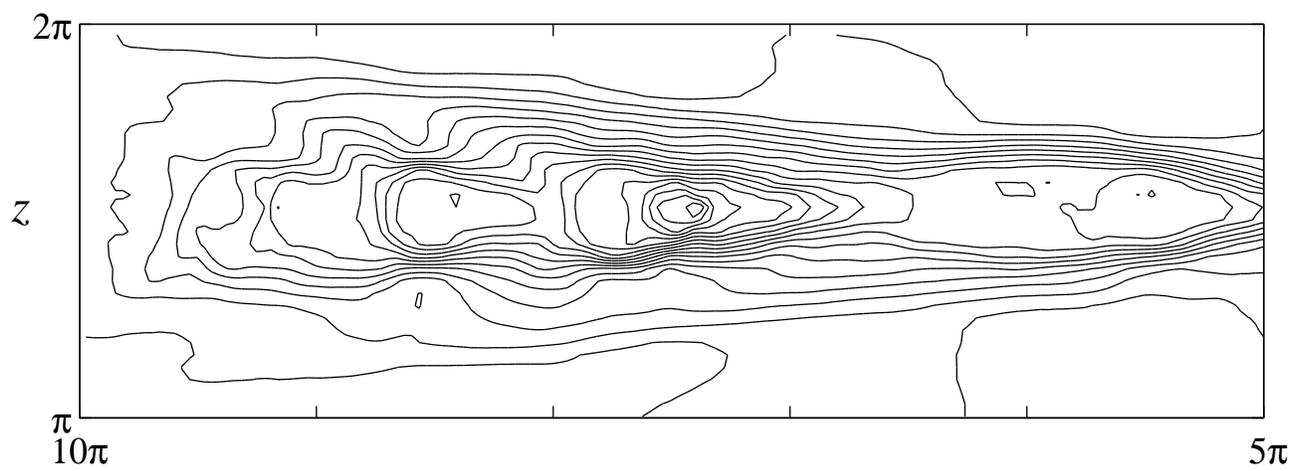
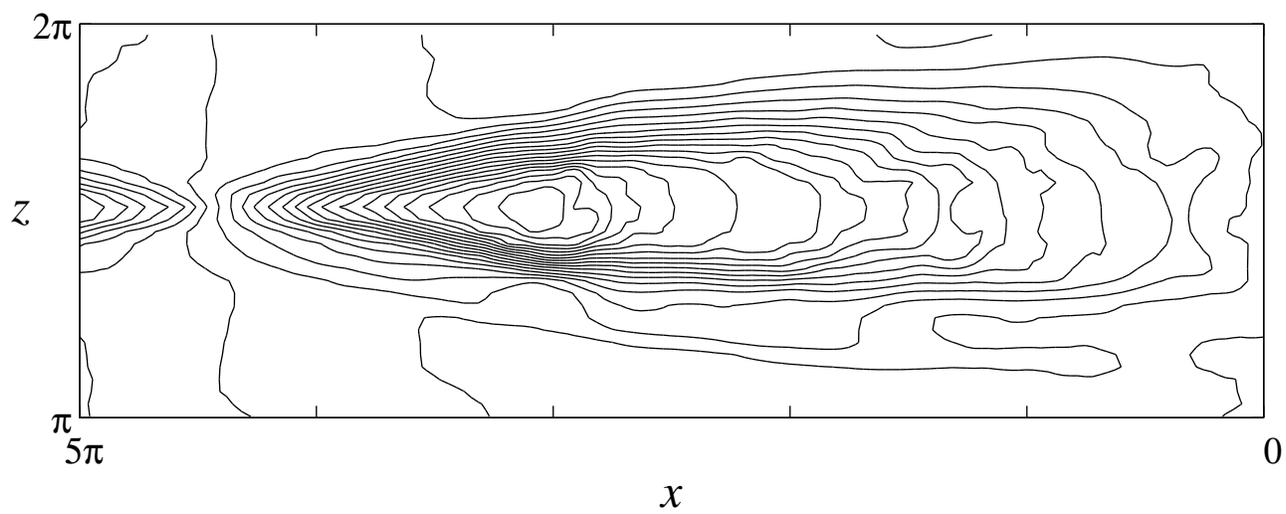
\end{document}